\documentclass[12pt]{article}

\usepackage{citesort}
\usepackage{amsthm}
\usepackage{epsfig}

\newenvironment{keywords}{\vskip1em%
       \noindent\small{\bf Key words.\ }\ignorespaces}
      {\par\normalsize}

\newtheorem{theorem}{Theorem}
\newtheorem{lemma}[theorem]{Lemma}

\newtheorem{fact}[theorem]{Fact}

\newcommand{\comment}[1]{}

\begin{document}

\title{
The Enhanced Double Digest Problem for DNA Physical Mapping}

\author{
Ming-Yang Kao\thanks{Department of Computer Science, Yale University,
New Haven, CT 06520, USA.  Email: kao-ming-yang@cs.yale.edu.  Research
supported in part by NSF Grant 9531028.}
\and
Jared Samet\thanks{Yale College, New Haven, CT 06520, USA. Email:
jared.samet@yale.edu.}
\and
Wing-Kin Sung\thanks{Department of Computer Science, Yale University,
New Haven, CT 06520, USA.  Email: sung-ken@cs.yale.edu.}}

\maketitle

\begin{abstract}
The {\it double digest problem} is a common NP-hard approach
to constructing physical maps of DNA sequences.
This paper presents a new approach called
the {\it enhanced double digest problem}.
Although this new problem is also NP-hard,
it can be solved in linear time in certain 
theoretically interesting cases.
\end{abstract}


\newcommand{\eA}{{\cal A}}
\newcommand{\eB}{{\cal B}}

\section{Introduction}

The physical mapping of DNA is a key problem in computational
biology \cite{Karp93}.
A {\it map} of a DNA sequence consists of
the locations of some given small sequences like e.g. GAATTC.
Biologists use such maps in a preparatory step to determine
the target DNA sequence \cite{NaSm75}.

A common technique of constructing maps
uses restriction enzymes to cut a DNA sequence at the positions
where a particular short DNA sequence appears.
These positions are called {\it restriction sites}.
One approach to modeling map construction is
the {\it double digest} (DD) problem.
Given two restriction enzymes $\eA$ and $\eB$,
this approach cuts
a target DNA sequence using enzyme $\eA$, enzyme $\eB$, and both
enzymes, separately.
It is a biology fact that the restriction sites for enzymes $\eA$ and $\eB$
do not coincide. Throughout this paper, we make use of this fact.
Let $A$, $B$ and $C$ be the three multisets of
the lengths of the fragments formed
after applying
enzyme ${\cal A}$, enzyme ${\cal B}$ and
both enzymes to the target DNA sequence, respectively.
Given $A$, $B$ and $C$, the DD problem asks for
permutations of the
lengths in $A$ and $B$ such that if these sets of lengths
are plotted on top of one another,
the lengths of all the resulting subintervals formed due to
overlapping match exactly
the lengths in $C$.
See Figure~\ref{fig-DD-example} for an example.

\begin{figure}[ht]
\begin{center}
\epsfig{figure=DDP-example.epsi}
\end{center}
\caption{Stripes (a), (b) and (c) show the fragments resulting from
the applications of
enzyme $\eA$, enzyme $\eB$ and both enzymes, respectively.
In strip (c), the subfragments are created due to the overlapping
between fragments in (a) and those in (b).}
\label{fig-DD-example}
\end{figure}

Many algorithms
\cite{Pevzner92,Pevzner95,SW91,WG86}
have been proposed for the DD problem.
Stefik \cite{Stefik78} gave the first algorithm
using artificial intelligence.
Fitch, Smith and Ralph \cite{FSR83}
reduced the DD problem to the set partition problem.
Goldstein and Waterman \cite{gw87} approached this
problem with
a stochastic annealing heuristic for the traveling salesman problem.
They also showed that
the DD problem is NP-hard by reducing
the set partition problem to it.

This paper suggests a new approach,
called the {\it enhanced double digest} (EDD) problem.
The EDD problem uses $A$, $B$, $C$ and some additional length information;
see Section~\ref{sec-definition} for the details of the approach.
Although the EDD problem is still NP-hard,
we show that if the lengths in $C$ are all distinct,
it can be solved in linear time.
We also generalize the algorithm for the case where
the number of duplicates in $C$ is bounded by a constant.
The time complexity of this generalized algorithm remains linear.

Section~\ref{sec-definition} details the new approach to
define the EDD problem formally.
Section~\ref{sec-linear} gives the linear-time algorithm
for the case where $C$ is duplicate-free.
Also, it generalizes the algorithm to handle
a small number of duplicate lengths.
Section~\ref{sec-np} proves that the EDD problem is NP-hard.
Section~\ref{sec-conclusion} concludes with some directions
for further work.

\newcommand{\frag}[1]{{\widehat{#1}}}

\section{Problem formulation} \label{sec-definition}

Consider a target DNA sequence and two restriction enzymes $\eA$ and $\eB$.
\begin{itemize}
\item By applying enzyme $\eA$ (respectively, $\eB$) to the target DNA sequence,
  we obtain $p$ (respectively, $q$) fragments.
  Let $A = \{a_1, \ldots, a_p\}$ (respectively, $B = \{b_1, \ldots, b_q\}$) be
  the multiset of the lengths of these $p$ (respectively, $q$) fragments.
\item For $i = 1, \ldots, p$,
  let $\frag{a}_i$ be the fragment corresponding to $a_i$.
  We apply enzyme $\eB$ to the fragment $\frag{a}_i$ and obtain
  a set of subfragments. Let $AB_i$ be the multiset of the lengths of
  these subfragments.
\item For $j = 1, \ldots, q$,
  let $\frag{b}_j$ be the fragment corresponding to $b_j$.
  We apply enzyme $\eA$ to the fragment $\frag{b}_j$ and obtain
  a set of subfragments. Let $BA_j$ be the multiset of the lengths of
  these subfragments.
\end{itemize}

For the example in Figure~\ref{fig-DD-example}, the following length
information is gathered:
\[
\begin{array}{ll}
\bullet & A = \{a_1 = 9, a_2 = 12, a_3 = 15, a_4 = 17, a_5 = 37\};
B = \{b_1 = 6, b_2 = 38, b_3 = 46\}; \\
\bullet & AB_1 = \{3, 6\}; AB_2 = \{12\}; AB_3 = \{15\}; AB_4 = \{17\};
  AB_5 = \{8, 29\}; \\
\bullet & BA_1 = \{6\}; BA_2 = \{3, 8, 12, 15\}; BA_3 = \{17, 29\}.
\end{array}
\]
It is easily verified that the data found in this way
has the following properties:
\begin{fact} \label{fact-G} \hspace*{1in}
\begin{enumerate}
\item For $i = 1, \ldots, p$, $a_i = \sum_{c \in AB_i} c$.
  For $j = 1, \ldots, q$, $b_j = \sum_{c \in BA_j} c$.
\item $\bigcup_i AB_i = \bigcup_j BA_j = C$.
\item $|C| = |A| + |B| - 1$.
\end{enumerate}
\end{fact}
\begin{proof}
Straightforward.
\end{proof}

Given $A, B, AB_1, \ldots, AB_p, BA_1, \ldots, BA_q$,
the {\it enhanced double digest problem} ${\cal P}$ asks for
a {\it valid permutation} $(\pi_A, \pi_B)$
of the elements in $A$ and $B$ such that the following can be achieved.
When the fragments $\frag{a}_i$ for $a_i \in A$ and
$\frag{b}_j$ for $b_j \in B$
are plotted on the same line
according to the order given by $\pi_A$ and $\pi_B$,
a set of subfragments is formed
due to overlapping.
The multiset $C$ of the lengths of these subfragments
is required to be equal to $\cup_{i=1}^p AB_i = \cup_{j=1}^q BA_j$.
In addition,
\begin{itemize}
\item for every $a_i \in A$ (respectively, $b_j \in B$),
$AB_i$ (respectively, $BA_j$) is equal to
the multiset of the lengths of the subfragments which
overlap with $\frag{a}_i$ (respectively, $\frag{b}_j$).
\end{itemize}

Note that an instance of this problem may have no solution or
more than one valid permutation.
The algorithms
given in Section~\ref{sec-linear} can recover all
valid permutations, if any exists.

\newcommand{\problem}{{\cal P}}

\section{An efficient algorithm}
\label{sec-linear}

Unless otherwise stated, this section assumes that $C$ has no duplicates.
Let $n = |C|$.
This section shows that the EDD problem $\problem$
can be solved in $O(n)$ time.

Section~\ref{sec-graph-represent}
formulates the EDD problem as a graph problem.
Section~\ref{sec-distinct} describes the linear-time
algorithm. Section~\ref{sec-general} discusses how to generalize this
linear-time algorithm to the case where $C$ may contain a small number of
duplicates.

\subsection{A graph representation} \label{sec-graph-represent}
Given $A, B, AB_1, \ldots, AB_p, BA_1, \ldots, BA_q$,
we construct an undirected graph $G$ as follows.
\begin{itemize}
\item The node set of $G = A \cup B \cup C$.
\item For every $a_i \in A$ and every $x \in C$, $(a_i, x) \in G$
  if $x \in AB_i$.
\item For every $b_j \in B$ and every $x \in C$,
  $(b_j, x) \in G$ if $x \in BA_j$.
\end{itemize}

\begin{figure}
\begin{center}
\epsfig{figure=exp-tree2.epsi}
\end{center}
\caption{The graph $G$ in (a) is constructed from the example
in Figure~\ref{fig-DD-example}.
$G$ can be redrawn into a tree as shown in (b).
The superscript $A, B$ or $C$ of each node denotes whether the node belongs to
$A, B$ or $C$.}
\label{fig-eg-tree}
\end{figure}

From the definition, we can observe that $G$ satisfies the following lemma.
\begin{lemma} \label{lem-simple-prop}
$G$ is connected.
For each node in $A \cup B$, its degree is at least $1$ and
it is adjacent to nodes in $C$ only.
Also, every node in $C$ connects to exactly one node in $A$ and one node in $B$.
\end{lemma}
\begin{proof}
Straightforward based on the assumption that $C$ has no duplicates.
\end{proof}

If $\problem$ has a valid permutation,
$G$ has two more properties as stated in Lemma~\ref{lem-property}.
Figure~\ref{fig-eg-tree} illustrates an example.
A {\it diameter} of a tree is a path with the largest number of edges.
A {\it dangler} is a $2$-node-long path.
Given a tree $T$, a subtree $\tau$ of $T$ is said to be {\it hanged on}
a path $P$ in $T$ if $\tau$ is a tree in the forest $T - P$.

\begin{lemma} \label{lem-property}
If $\problem$ has a valid permutation, then the following statements hold.
\begin{enumerate}
\item \label{item2}
$G$ is a tree.
\item \label{item3}
For any diameter $S$ of $G$,
the subtrees hanged on $S$ must be danglers.
\end{enumerate}
\end{lemma}
\begin{proof} \hspace{1in}

Statement \ref{item2}.
To prove by contradiction,
suppose that $G$ contains a cycle $D$. By the construction of $G$,
$D$ must be of the form
\[a_{i_1}, c_{k_1}, b_{j_1}, c_{k_2}, a_{i_2}, c_{k_3}, b_{j_2}, c_{k_4},
 \ldots, c_{k_{2z}}, a_{i_{z+1}},\]
where $i_1 = i_{z+1}$;
$a_{i_1}, \ldots, a_{i_z} \in A$; $b_{j_1}, \ldots, b_{j_z} \in B$; and
$c_{k_1}, \ldots, c_{k_{2z}} \in C$.

By definition, if $a_{i}, c_{k}, b_{j}$ is a path in $G$,
then $\frag{a}_i$ and $\frag{b}_j$ overlap by $\frag{c}_{k}$ in any
valid permutation of $\problem$.
Thus, for $1 \leq \ell \leq z-1$,
the existence of the subpath $a_{i_{\ell}}, \ldots, a_{i_{\ell + 2}}$ of $D$
in $G$ means that
$\frag{b}_{i_{\ell}}$ overlaps
with $\frag{a}_{i_{\ell}}$ and $\frag{a}_{i_{\ell+1}}$
and $\frag{b}_{i_{\ell+1}}$ overlaps with
$\frag{a}_{i_{\ell+1}}$ and $\frag{a}_{i_{\ell+2}}$.
To enable both $\frag{b}_{i_{\ell}}$ and $\frag{b}_{i_{\ell+1}}$
overlap with $\frag{a}_{i_{\ell+1}}$,
$\frag{a}_{i_{\ell+1}}$ must be in the middle of $\frag{a}_{i_{\ell}}$
and $\frag{a}_{i_{\ell+2}}$ for $1 \leq \ell \leq z-1$.
Consequently, for $1 \leq \ell \leq z-1$,
$\frag{a}_{i_{\ell}}$ is in the middle of $\frag{a}_{i_1}$ and
$\frag{a}_{i_{z+1}} = \frag{a}_{i_1}$,
which is impossible.

Statement \ref{item3}.
For any diameter $S$ of $G$,
we show that every subtree $\tau$ hanged on $S$ must
be a dangler.
First, $\tau$ must be hanged on $S$ at a node in $A \cup B$.
Otherwise, if $\tau$ is hanged on $S$ at a node $c \in C$,
$c$ has degree greater than
$2$, contradicting Lemma~\ref{lem-simple-prop}.
Then, $\tau$ has more than one node because
the root of $\tau$ is a node in $C$ and must be of degree $2$.
If $\tau$ cannot have
more than $2$ nodes, Statement~\ref{item3} follows.

\begin{figure}
\begin{center}
\epsfig{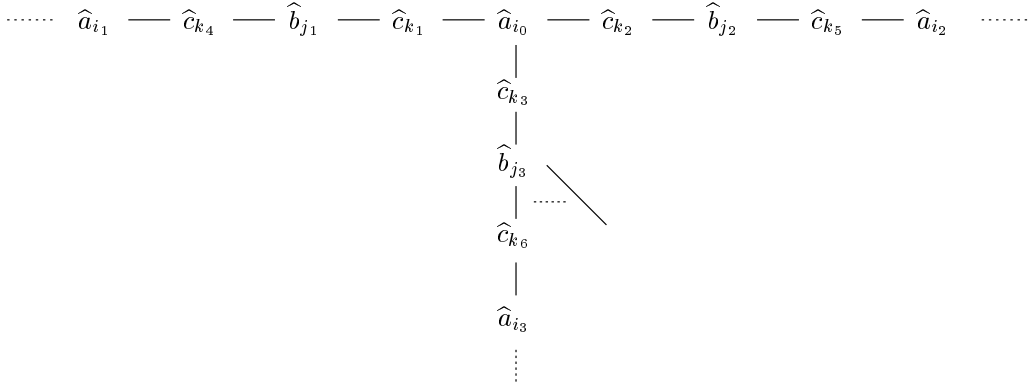}
\end{center}
\caption{In this example, all $a_i \in A$, $b_j \in B$ and $c_k \in C$.}
\label{fig-invalid-eg}
\end{figure}

To prove by contradiction,
suppose that $\tau$ has more than two nodes.
Without lost of generality, assume that $\tau$ is hanged on $S$ at a node
$a_{i_0} \in A$ and the root of $\tau$ is a node $c_{k_3} \in C$.
Note that $c_{k_3}$ has another neighbour, say $b_{j_3}$, from $B$.
If $\tau$ contains more than two nodes,
$b_{j_3}$ must has a child, say $c_{k_6}$, from $C$ and
$c_{k_6}$ must has a child, say $a_{k_3}$, from $A$.
Thus, $\tau$ must have a root-to-leaf path of length more than $4$.
Then, the two paths from $a_{i_0}$ to both ends of $S$
must be of length more than $4$.
Otherwise, $S$ cannot be a diameter of $G$.
From those observations, $G$ has the pattern
shown in Figure~\ref{fig-invalid-eg}.
According to the pattern, $\frag{b}_{j_1}, \frag{b}_{j_2}$ and $\frag{b}_{j_3}$
overlap with $\frag{a}_{i_0}$.
Therefore, in any valid permutation,
one of $\frag{b}_{j_1}, \frag{b}_{j_2}$ and $\frag{b}_{j_3}$,
say $\frag{b}_{j_2}$, must be in the middle of the other two fragments
and $\frag{b}_{j_2}$ can only overlap with $\frag{a}_{i_0}$.
However, according to the pattern in Figure~\ref{fig-invalid-eg},
for $\ell = 1, 2, 3$, $\frag{b}_{j_{\ell}}$ overlaps
with another fragment $\frag{a}_{i_{\ell}}$,
reaching a contradiction.
\end{proof}

Now, we know that if $\problem$ has a valid permutation, $G$ satisfies the
two properties of Lemma~\ref{lem-property}.
The remainder of this section show that the converse of this statement is
also true.
Suppose that $G$ is a tree with a diameter $S$ such that all
the subtrees hanged on $S$ are danglers.
We define $\pi_C$ to be a permutation on $C$ formed
by a search defined below.

\begin{description}
\item[{\it Dangler-first search:}]
Traverse $G$ starting from one end of $S$ to the other end of $S$;
read off the nodes in $C$ on $S$; whenever meet any node $x$ with degree
greater than $2$,
read off the nodes in $C$ in the danglers hanged on $S$ at $x$ in any order
and continue to traverse $S$.
\end{description}

\begin{lemma} \label{lem-read-c}
The elements in each $AB_i$ form
a consecutive subsequence in $\pi_C$.
Similarly, the elements in each $BA_j$ form
a consecutive subsequence in $\pi_C$.
\end{lemma}
\begin{proof}
For each $i$, if $AB_i$ contains only one element, then the lemma follows.
Otherwise, $a_i$ is of degree at least $2$. 
Then, $a_i$
must be on the diameter $S$.
Let  $c$ and $c'$ be elements in $AB_i$
which are the two neighbours of $a_i$ on $S$.
The remaining nodes in $AB_i$
must be located in the danglers hanged on $S$ at $a_i$.
By dangler-first search, all the elements in $AB_i$
must form a consecutive subsequence in $\pi_C$.
By symmetry, for each $j$, the elements in $BA_j$
must form a consecutive subsequence in $\pi_C$.
\end{proof}

By Lemma~\ref{lem-read-c}, $\pi_C$ can be partitioned into $p$ subintervals
such that the $r$th interval contains the elements
in $AB_{i_r}$ for $r = 1, \ldots, p$. Let $\pi_A$ be the permutation
$(a_{i_1}, \ldots, a_{i_p})$.
Similarly, $\pi_C$ can be partitioned into $q$ intervals
such that the $s$th interval contains the elements in $BA_{j_s}$
for $s = 1, \ldots, q$.
Let $\pi_B$ be the permutation $(b_{j_1}, \ldots, b_{j_q})$.
We call $(\pi_A, \pi_B)$ the {\it induced permutation} of $\pi_C$.

\begin{lemma} \label{lem-valid-solution}
The induced permutation $(\pi_A, \pi_B)$ of $\pi_C$ is a valid permutation
of $\problem$.
\end{lemma}
\begin{proof}
Suppose the lengths from $A, B$ and $C$ are plotted
on the same line according to the order given by $\pi_A$, $\pi_B$
and $\pi_C$, respectively.
Consider the stripes formed from $A$ and $C$.
By Fact~\ref{fact-G} and Lemma~\ref{lem-read-c},
for each $i$, $\frag{a}_i$ overlaps with $\frag{c}$ for
all $c \in AB_i$.
By symmetry, for each $j$, $\frag{b}_j$ overlaps
with $\frag{c}$ for all $c \in BA_j$.
Then, by the definition of the EDD problem,
$(\pi_A, \pi_B)$ is a valid permutation.
\end{proof}

\begin{theorem} \label{corol-summary}
Given the enhanced double digest problem $\problem$
and its corresponding graph $G$,
$\problem$ has a valid permutation if and only if
$G$ satisfies the two properties in Lemma~\ref{lem-property}.
\end{theorem}
\begin{proof}
The only-if part follows from
Lemma~\ref{lem-property}.
The if part follows from Lemma~\ref{lem-valid-solution}.
\end{proof}

\subsection{A linear-time algorithm for a duplicate-free $C$}
\label{sec-distinct}

This section describes how to
compute a valid permutation of $\problem$
in $O(n)$ time.
The algorithm is as follows.

{\bf Algorithm} Enhanced-Double-Digest
\begin{enumerate}
\setlength{\itemsep}{-\parsep}
\item Construct the graph $G$ corresponding to
  $\problem$.
\item If $G$ does not satisfy
  the two properties in Lemma~\ref{lem-property},
  then
  return ``no valid permutation''.
\item Find the permutation $\pi_C$ using dangler-first search.
\item Find the induced permutation $(\pi_A, \pi_B)$ of $\pi_C$.
\item Return $(\pi_A, \pi_B)$.
\end{enumerate}

\begin{lemma}
Algorithm Enhanced-Double-Digest
can correctly find a valid permutation in $O(n)$ time.
\end{lemma}
\begin{proof}
First,
by Lemma~\ref{lem-valid-solution} and Theorem~\ref{corol-summary},
Enhanced-Double-Digest is correct.
As for its time complexity,
Step 1 requires $O(n)$ time as $G$ contains $2n$ edges
and we can find each edge in $O(1)$ time.
Step 2 checks whether $G$ satisfies
the two properties in Lemma~\ref{lem-property}.
For property \ref{item2}, we can determine whether
a graph is a tree in $O(n)$ time.
For property \ref{item3},
we can compute a diameter of a tree in linear time first,
then, we verify whether $G$ satisfies property \ref{item3}
by detecting whether the subtrees hanged on the diameter are
danglers.
Thus, Step 2 requires $O(n)$ time.
Step 3 finds $\pi_C$ using dangler-first search.
Since the search scans every node in $G$ once, it runs in $O(n)$ time.
Step 4 finds
the induced permutation $(\pi_A, \pi_B)$ of $\pi_C$ in $O(n)$ time.
In summary, a valid permutation of $\problem$
can be computed in $O(n)$ time.
\end{proof}

By modifying Algorithm Enhanced-Double-Digest slightly, we can report all valid
permutations of $\problem$.
First, observe that the valid permutations of $\problem$
depend on the possible permutations $\pi_C$. There are three cases.

Case 1: $G$ does not have any dangler. Then, there is a unique $\pi_C$.
Thus, the current algorithm reports all valid permutations of $\problem$.

Case 2: $G$ has one set of danglers hanged on one node of its diameter. Then,
the possible permutations $\pi_C$ depend on the permutation of the set of nodes in
the danglers which belong to $C$.
For the example in Figure~\ref{fig-eg-tree},
the possible permutations $\pi_C$ can be represented by
\[6, 3, \mbox{permutation}(12, 15), 8, 29, 17.\]
All valid permutations $\pi_A$ and $\pi_B$ can be represented by
$9, \mbox{permutation}(12, 15), 37, 17$ and
$6, 38, 46$, respectively.
These valid permutations can be reported by modifying Steps 3 and 4
of the algorithm.
The time complexity of the modified algorithm is still $O(n)$.

Case 3: $G$ has $k$ sets of danglers hanged on $k$ respective nodes of its diameter.
Then, the possible permutations $\pi_C$ can be represented similarly, except
that each $\pi_C$ contains $k$ permutation blocks.
The above modified algorithm is sufficient to report
all valid permutations of $\problem$.

\subsection{A general algorithm for $C$ with few duplicates}
\label{sec-general}

The algorithm Enchanced-Double-Digest in Section~\ref{sec-distinct} can solve
the EDD problem if $C$ contains no duplicates.
Here, we give an algorithm which works without this assumption. 

First, we consider the following example.

\[ \begin{array}{ll}
\bullet & A = \{a_1 = 18, a_2 = 19\};
B = \{b_1 = 4, b_2 = 5, b_3 = 7, b_4 = 8, b_5 = 13\}; \\
\bullet & AB_1 = \{5, 6, 7\}; AB_2 = \{4, 7, 8\}; \\
\bullet & BA_1 = \{4\}; BA_2 = \{5\}; BA_3 = \{7\}; BA_4 = \{8\}; BA_5 = \{6, 7\}.
\end{array} \]

\begin{figure}
\begin{center}
\epsfig{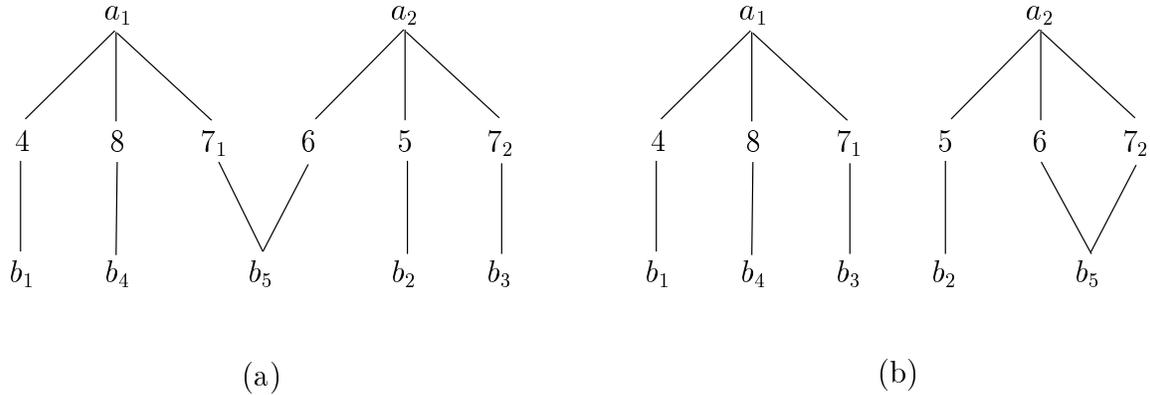}
\end{center}
\caption{(a) is the case where $7_1 \in BA_5$ and $7_2 \in BA_3$;
(b) is the case where $7_1 \in BA_3$ and $7_2 \in BA_5$. }
\label{fig-general-eg}
\end{figure}
In this example, there are two $7$'s in $C = \cup_i AB_i = \cup_j BA_j$.
These two $7$'s in fact represent two
different subfragments in the target DNA sequence. To distinguish them,
let the copy of $7$ in $AB_1$ be $7_1$ and that in $AB_2$ be $7_2$.
Since $7$ also belongs to $BA_3$ and $BA_5$,
there are two possible combinations, namely,
(a) $7_1 \in BA_5$ and $7_2 \in BA_3$ and
(b) $7_1 \in BA_3$ and $7_2 \in BA_5$.
Figure~\ref{fig-general-eg}(a) and \ref{fig-general-eg}(b)
illustrate the graph $G$ for both cases; from these two graphs $G$,
we can obtain a valid permutation from combination (a).
Therefore, we can handle duplicates in $C$ by giving them different
subscripts. Then, all the elements in $C$ are different and
we can solve the enhanced double digest problem using
the algorithm Enhanced-Double-Digest in Section~\ref{sec-distinct}.
More precisely, we have the following algorithm.

\begin{enumerate}
\setlength{\itemsep}{-\parsep}
\item If $C$ contains duplicates, then
 we assign a unique subscript to each duplicate.
\item For each possible combinations of the subscripts in the duplicates, we
execute Enhanced-Double-Digest to compute a valid permutation.
\end{enumerate}

Let $\ell$ be the number of duplicates in $C$. The above algorithm execute
Enhanced-Double-Digest for at most $\ell!$ time.
Therefore, a valid permutation can be
computed in $O(\ell! n)$ time.
Thus, if $\ell$ is constant,
the generalized algorithm still runs in linear time.

\section{The enhanced double digest problem is NP-hard} \label{sec-np}
This section proves the NP-hardness of
the enhanced double digest problem
by a reduction from the Hamiltonian Path problem \cite{GJ79}.

Given an undirected graph $H$, we show that in polynomial time,
we can construct an EDD instance ${\cal Q}$ so that $H$ contains a
hamiltonian path if and only if ${\cal Q}$ has a valid permutation. 
For ease of prove, we augment $H$ with two new nodes $t$ and $z$.
All nodes originally in $H$ have edges to $t$.
In addition, we add an edge $(t, z)$ to $H$.
Note that the original $H$ contains a hamiltonian path
if and only if the amended $H$
has a hamiltonian path.
Let $\ell$ be the number of nodes in $H$.
Assume that the nodes in $H$ are labeled by
$\{1, 2, \ldots, \ell\}$.
For each node $v$,
let $\kappa(v)$ be the number of neighbours of $v$.
Let $v' = v + \ell$.
The EDD instance ${\cal Q}$ is given
the following length information.
Note that this length information can be constructed
from $H$ in polynomial time.
\begin{itemize}
\item $A = \{ a_v \mid v \in H \}$ where
  $a_z = t'$, $a_t = t + \sum_{u \in H - \{t, z\}} u'$
  and $a_v = v + \sum_{(u, v) \in H} u'$ for $v \neq z, t$. 
  Also, $AB_z = \{t'\}$;
  $AB_t = \{u' \mid u \in H - \{t, z\}\} \cup \{t\}$; and
  $AB_v = \{u' \mid (u, v) \in H\} \cup \{v\}$ for $v \neq z$.
\item $B = \{ b_v, b_{v(1)}, \ldots, b_{v(\kappa(v)-1)} \mid v \in H-\{z\} \}$
  where $b_v = v+v'$ and $b_{v(i)} = v'$ for all $v \in H-\{z\}$ and
  all $i \leq \kappa(v)-1$.
  Also, $BA_v = \{v, v'\}$ and $BA_{v(i)} = \{v'\}$.
\end{itemize}

\begin{lemma}
$H$ has a hamiltonian path if and only if 
there is a valid permutation for ${\cal Q}$.
\end{lemma}
\begin{proof}
The two directions are proved as follows.

\begin{figure}[ht]
\begin{center}
\epsfig{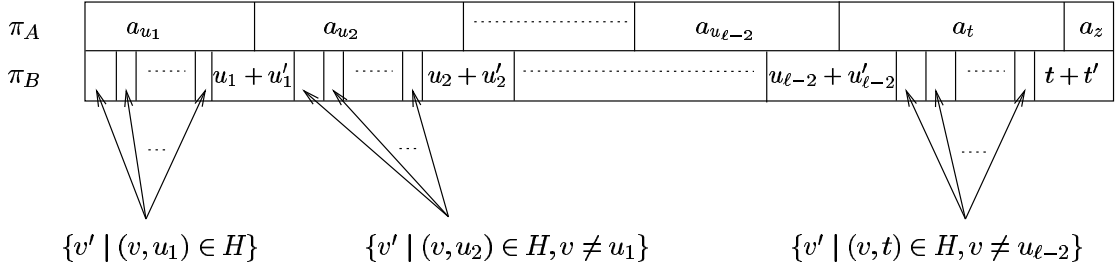}
\end{center}
\caption{The permutations $\pi_A$ and $\pi_B$ of $A$ and $B$, respectively.}
\label{fig-permut}
\end{figure}
($\Longrightarrow$) 
Let $u_1, u_2, \ldots, u_{\ell-2}, t, z$ be a hamiltonian path in $H$.
Let $\pi_A$ and $\pi_B$ be permutations of $A$ and $B$
as shown in Figure~\ref{fig-permut}.
It is easy to check that
$(\pi_A, \pi_B)$ is a valid permutation to ${\cal Q}$.

($\Longleftarrow$)
Let $(\pi_A, \pi_B)$ be a valid permutation of ${\cal Q}$.
The remainder of this proof shows that the ordering of the lengths in $\pi_A$
defines a hamiltonian path in $H$.

Assume the lengths from $A$ are plotted on a line according to the
order given by $\pi_A$ and similarly, the lengths from $B$ are also
plotted on this line according to $\pi_B$.
For each $v \in H$, the line fragment corresponds to $a_v \in A$
is called $\frag{a}_v$.
For each $v \in H - \{z\}$,
the line fragment corresponds to $b_v \in B$, is called $\frag{b}_v$.

For every $v \in H - \{z\}$,
since $BA_{v} = \{v, v'\}$,
$\frag{b}_v$ overlaps with two consecutive line fragments from $A$;
in addition,
the overlapping regions
between $\frag{b}_v$ and these two line fragments
must be of length $v$ and $v'$, respectively.
Observe that $v \in AB_v$ and $v \not\in AB_u$ for all $u \neq v$.
One of these two fragments,
which overlaps with $\frag{b}_v$, must be $\frag{a}_v$.
The other line fragment can be $\frag{a}_u$ for any $u \in H$
with $v' \in AB_u$, i.e., $(v, u) \in H$.

Let $\pi_A = (a_{u_1}, \ldots, a_{u_l})$.
>From the above argument, we know that, for every two consecutive
line fragments $\frag{a}_i$ and $\frag{a}_{i+1}$,
there exists a fragment $\frag{b}_v$ (where $v$ is either $u_i$ or
$u_{i+1}$) which overlaps with both $\frag{a}_{u_i}$ and $\frag{a}_{u_{i+1}}$.
The above argument also implies that $(u_i, u_{i+1}) \in H$.
Thus, $u_1, \ldots, u_{\ell}$ forms a path in $H$.
As $u_1, \ldots, u_{\ell}$ contains all the $\ell$ nodes of $H$,
this path is a hamiltonian path.
\end{proof}

\section{Further research directions} \label{sec-conclusion}

This highly theoretical work can be extended in several directions.
One direction is to design a series of laboratory procedures that
can actually produce the input length information in the required form.
Another direction is to consider
the problem of more than $2$ digesting enzymes. Using multiple
enzymes could help resolve the issue of multiple solutions
that arise when there are danglers or duplicate subfragment lengths.
Also, the extra input may actually make the problem solvable in
a shorter period of time.
The third direction is to have a probabilistic analysis of
the number of duplicates in $C$, when the length of the target DNA sequence
is given. 
It would be the most meaningful to conduct such analysis under a
probabilistic model that is derived specifically for feasible
laboratory procedures.
Lastly, this paper does not address the issue of noise in the length data.
From the practical point of view, handling noise effectively is very
important.

\section{Acknowledgments}
We wish to thank the anonymous referees for many helpful suggestions.


\end{document}